\documentclass[aps,prl,preprint,superscriptaddress,footinbib,amsmath,amssymb]{revtex4-1}

\usepackage[T1]{fontenc}
\usepackage[utf8]{inputenc}
\usepackage{lmodern}
\usepackage{textcomp}
\usepackage{microtype}
\usepackage{graphicx}
\usepackage{hyperref}

\usepackage{bm}

\begin{document}

\title{Polarized photoluminescence clocks ultrafast pseudospin relaxation in graphene}

\author{Thomas Danz}
\author{Andreas Neff}
\altaffiliation[Present address: ]{Leibniz Institute of Surface Modification, Leipzig, Germany}
\affiliation{4th Physical Institute -- Solids and Nanostructures, University of G\"ottingen, G\"ottingen, Germany}
\affiliation{These authors contributed equally to this work.}
\author{John H. Gaida}
\author{Reiner Bormann}
\author{Claus Ropers}
\author{Sascha Sch\"afer}
\email{schaefer@ph4.physik.uni-goettingen.de}
\affiliation{4th Physical Institute -- Solids and Nanostructures, University of G\"ottingen, G\"ottingen, Germany}

\date{\today}

\begin{abstract}
Electronic states in 2D materials can exhibit pseudospin degrees of freedom, which allow for unique carrier-field interaction scenarios. Here, we investigate ultrafast sublattice pseudospin relaxation in graphene by means of polarization-resolved photoluminescence spectroscopy. Comparison with microscopic Boltzmann simulations allows to determine a lifetime of the optically aligned pseudospin distribution of $12\pm 2\,\text{fs}$. This experimental approach extends the toolbox of graphene pseudospintronics, providing novel means to investigate pseudospin dynamics in active devices or under external fields.
\end{abstract}

\maketitle

Present-day electronic devices process, store, and transport information based on charge carriers. Recent developments in spintronics \cite{Wolf2001,Zutic2004,Han2014} have extended these capabilities by additionally making use of the electronic spin. Moreover, depending on the local environment, carriers can be equipped with additional pseudospin degrees of freedom, including sublattice, valley, and layer pseudospin, which may be exploited in future information technology. These angular momentum components exhibit rich physical phenomena, not unlike the ones observed for the intrinsic spin of electrons \cite{Mecklenburg2011,Song2015,Xu2014,Schaibley2016}.

For example, the valley pseudospin is exploited in valleytronics \cite{Schaibley2016} by manipulating the occupation of degenerate but inequivalent chiral electron states. In particular, in hexagonal 2D materials such as transition metal dichalcogenides (TMDCs), selection rules enable a direct optical manipulation of the carrier pseudospin \cite{Cao2012,Zeng2012,Xu2014,Mak2014,Schaibley2016}. Manifestations of valley pseudospin polarization in TMDCs include the valley-spin analogue of the spin Hall effect \cite{Xiao2012,Mak2014}, and the valley Zeeman effect \cite{Srivastava2015}.

Even single-layer graphene as the most simple 2D material displays valley and sublattice pseudospin \cite{Geim2007}. While, similar to the case of TMDCs, the valley pseudospin distinguishes the occupation within the Dirac cones at the $K$ and $K'$ points, the sublattice pseudospin controls the relative phase of the electron wave function on the two hexagonal sublattices \cite{Pesin2012}. In carrier momentum space, sublattice pseudospin is collinear with the carrier momentum relative to the Dirac point \cite{Geim2007,Mecklenburg2011}. Sublattice pseudospin conservation enables Klein tunneling in graphene, i.\,e. the counter-intuitive carrier transmission through infinitely high potential barriers \cite{Klein1929,Katsnelson2006,Young2009}. 

In graphene, coupling between carriers and optical fields \cite{Nair2008} is governed by sublattice pseudospin selection rules \cite{Mecklenburg2010}, so that the initial carrier populations created by linearly polarized interband excitation exhibit a pronounced angularly asymmetric population within the Dirac cones \cite{Malic2011}. However, intrinsic carrier scattering mechanisms are expected to rapidly destroy the optically imprinted pseudospin alignment. Ultrafast pseudospin relaxation has been addressed in a series of polarization-resolved transient optical spectroscopy experiments \cite{Chen2014,Mittendorff2014,Yan2014,Trushin2015,Yao2015,Yao2015a,Konig-Otto2016,Yan2017}, elucidating the role of carrier momentum isotropization in carrier thermalization and cooling processes \cite{Dawlaty2008,George2008,Sun2008,Newson2009,Breusing2011,Winnerl2011,Tani2012,Brida2013,Malard2013,Winnerl2013}. Although it is challenging to observe the fastest relaxation dynamics, most of the previous investigations suggest that sublattice pseudospin relaxation is complete within $50\,\text{fs}$ to $150\,\text{fs}$ \cite{Mittendorff2014,Yan2014,Trushin2015}. Theoretical models predict even faster dynamics \cite{Malic2012}, and a precise relation between energetic and momentum relaxation timescales has not been established experimentally, yet.

Here, we present evidence for a transient sublattice pseudospin state in graphene, obtained by analyzing the polarization-resolved photoluminescence (PL) from optically induced non-equilibrium carrier populations \cite{Lui2010,Liu2010,Stohr2010,Koyama2013}. We demonstrate that the initial pseudospin alignment is lost within $12\pm 2\,\text{fs}$ by comparing the experimentally observed optical polarization degree of the PL with microscopic simulations of Boltzmann rate equations. The results highlight the importance of efficient pseudospin relaxation in graphene, proceeding equally fast as energetic carrier thermalization, and an order of magnitude faster than the excitation of strongly coupled optical phonons \cite{Kampfrath2005}.

In our experiments, single-layer graphene on a sapphire substrate is optically excited using ultrashort laser pulses ($18\,\text{fs}$ pulse duration, $1.55\,\text{eV}$ photon energy, $80\,\text{MHz}$ repetition rate). The resulting blue-shifted PL is detected in a polarization-resolved manner by a grating spectrometer and a liquid nitrogen-cooled charge-coupled device (CCD). At the excitation conditions employed, no optically induced sample damage was observed. Special care was taken to ensure accurate calibration of the spectral and polarization responses of the detection system (see Supplemental Material \footnote{See Supplemental Material at [url], which includes additional Refs. \onlinecite{Li2009,Gulde2014,Mattevi2011,Wang2012,Ferrari2006,Malard2009,Ziman1960,Saito1998,Gruneis2003,Fermi1953,Moskova1994}, for a detailed description of the experimental setup, of the sample preparation procedure, and of the theoretical fundamentals of the microscopic simulations.}).

\begin{figure*}[t]
		\includegraphics[width=\textwidth]{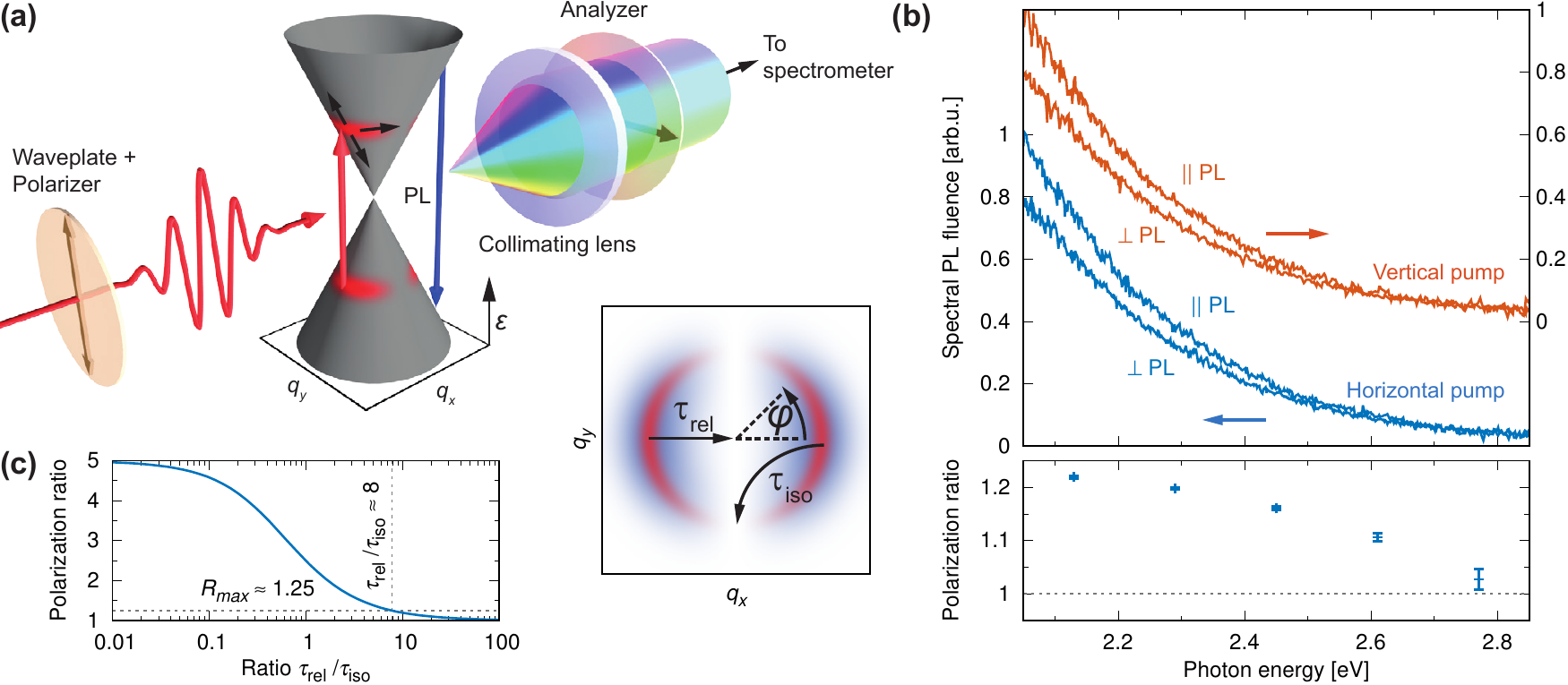}
    \caption{\label{fig1}
		\textbf{(a)} Schematics of experimental setup and PL emission process. Horizontally or vertically polarized ultrashort pump pulses trigger carrier dynamics around the Fermi energy of graphene. Resulting hot-carrier PL is collimated and recorded as a function of photon energy and polarization (horizontal or vertical).
		\textbf{Inset:} Projection of the initial (red) and broadened (blue) momentum distribution after optical excitation onto the $\bm q$ plane (momentum $\bm q$ relative to the $K$ and $K'$ points). Phenomenological relaxation timescales $\tau_\text{rel}$ and $\tau_\text{iso}$ are introduced.
		\textbf{(b) Top:} Representative spectra of parallel PL ($\parallel$ PL) and perpendicular PL ($\perp$ PL) for horizontal and vertical pump polarizations. The anisotropic nature of the PL emission is readily visible ($117\,\text{\textmu J}/\text{cm}^2$ incident fluence).
		\textbf{Bottom:} Polarization ratios extracted from the PL spectra above (averaged over both pump polarizations, $0.16\,\text{eV}$ bin width).
		\textbf{(c)} Polarization ratio as a function of $\tau_\text{rel}/\tau_\text{iso}$ in a model of two coupled, exponentially decaying carrier populations $\parallel$ and $\perp$. The indicated value of $R_\text{max}\approx 1.25$ is the maximum polarization ratio observed in the experiment according to the data in (b).
		}
\end{figure*}

Figure~\ref{fig1}a schematically depicts the experimental setup and the microscopic processes underlying PL emission. Ultrashort, linearly polarized optical pump pulses excite carriers from the valence into the conduction band of graphene (red arrow). Initially, the non-equilibrium carrier distribution is strongly anisotropic within each Dirac cone of the graphene band structure \cite{Malic2012}. During relaxation, carrier-carrier and carrier-phonon scattering cause energetic broadening of the carrier distribution and a loss of momentum anisotropy (black arrows). The emerging high-energy tail of the carrier distribution gives rise to a blue-shifted component of the PL by electron-hole recombination (blue arrow), as previously observed \cite{Lui2010,Liu2010,Stohr2010}, with recombining carriers at an energy $\pm\varepsilon$ emitting a PL photon of energy $E=2|\varepsilon|$. The carrier and phonon systems jointly thermalize due to an efficient energy transfer between carriers and a set of high-energy, strongly coupled optical phonons (SCOPs) on a hundred femtosecond timescale \cite{Kampfrath2005,Breusing2011}. The SCOPs in turn decay on a timescale on the order of a picosecond \cite{Bonini2007,Kang2010,Schafer2011,Chatelain2014}.

Energetic relaxation and azimuthal momentum randomization of the broadened carrier distribution can be characterized by two phenomenological relaxation times $\tau_\text{rel}$ and $\tau_\text{iso}$, respectively (see inset in Fig.~\ref{fig1}a). Our experiments yield access to momentum and energy relaxation rates, as the degree of PL polarization is governed by the ratio of these time constants. Specifically, in the limiting case of $\tau_\text{rel}\gg\tau_\text{iso}$, completely unpolarized PL emission is expected, whereas $\tau_\text{rel}\lesssim\tau_\text{iso}$ would result in a PL preferentially polarized parallel to the pump.

Representative PL spectra obtained in our experiments are depicted in Fig.~\ref{fig1}b. The PL intensity decreases mo\-not\-o\-nous\-ly over the observed photon energy range in a spectral shape that was previously described by black-body radiation \cite{Lui2010}. Most importantly, the intensity of the PL polarized parallel to the pump (indicated by $\parallel$) is up to $20\,\text{\%}$ higher than the intensity of the perpendicular PL component (indicated by $\perp$), evidencing considerable azimuthal carrier anisotropy within the PL lifetime. The agreement of the spectra for both horizontally (blue curves) and vertically (red curves) polarized pump beams demonstrates a proper polarization-dependent calibration of the detection. In order to quantify the polarization properties of the PL, we employ a polarization ratio
\begin{equation}
    R(E)=\frac{I_\parallel(E)}{I_\perp(E)}
\end{equation}
of the temporally integrated PL flux, $I_\parallel$ and $I_\perp$, polarized parallel and perpendicular to the pump, respectively, as a function of the photon energy $E$. Unpolarized PL emission corresponds to $R=1$, while fully polarized PL emission is represented by $R\rightarrow\infty$.

To gain a first phenomenological description of the expected PL polarization, we describe the azimuthal and energetic carrier relaxation in a simplified model by two coupled, exponentially decaying components $f_{c,\parallel}$ and $f_{c,\perp}$ of the relevant high-energy carrier population in the conduction band:
\begin{align}
    \frac{\text{d}f_{c,\parallel}}{\text{d}t}&=-\frac{f_{c,\parallel}}{\tau_\text{rel}}-\frac{f_{c,\parallel}-f_{c,\perp}}{\tau_\text{iso}},\label{eq:rates_par}\\
    \frac{\text{d}f_{c,\perp}}{\text{d}t}&=-\frac{f_{c,\perp}}{\tau_\text{rel}}-\frac{f_{c,\perp}-f_{c,\parallel}}{\tau_\text{iso}}.\label{eq:rates_perp}
\end{align}
In the low-energy optical regime and at a negligible doping level, the carrier occupation in the valence band is given by $f_v=1-f_c$ at any time. For both components, we adopt an azimuthal shape following the angular dependence of the off-axis carrier-field coupling matrix elements (for linear polarization) within the Dirac cones, i.\,e. the total carrier distribution is given by $f_c(\varphi,t)=f_{c,\parallel}(t)\,\sin^2\varphi+f_{c,\perp}(t)\,\cos^2\varphi$ (see inset in Fig.~\ref{fig1}a for the definition of $\varphi$) \cite{Malic2011}. Thereby, considering the limiting case of optical excitation pulses much shorter than the relevant relaxation times, the perpendicular component $f_{c,\perp}$ of the carrier distribution is initially zero, and is only populated subsequently due to the azimuthal momentum relaxation term.

Taking into account the azimuthal momentum distribution of both components, the temporally integrated PL flux and polarization ratio are deduced from the evolution of the carrier distribution \cite{Mecklenburg2010,Winzer2015} (see Supplemental Material \cite{Note1}). In this model, the polarization ratio is only a function of $\tau_\text{rel}/\tau_\text{iso}$, and an upper limit of the polarization ratio is given by $R=5$ (as shown in Fig.~\ref{fig1}c), as already the initial $f_{c,\parallel}$ population yields a small perpendicularly polarized PL flux. The experimentally observed maximum polarization ratio of $R\approx 1.25$ (cf. Fig.~\ref{fig1}b) is obtained at $\tau_\text{rel}/\tau_\text{iso}\approx 8$.
 
In order to quantitatively analyze the experimental data, we combine the rate equations introduced above with a microscopic carrier scattering model. Specifically, in Eqs.~\ref{eq:rates_par} and~\ref{eq:rates_perp}, we consider energy-dependent carrier distribution functions $f_{c,\parallel}(\varepsilon, t)$ and $f_{c,\perp}(\varepsilon, t)$, and replace the phenomenological exponential decay (described by $\tau_\text{rel}$) by Boltzmann rate equations, which include optical pulse absorption \cite{Mecklenburg2010,Malic2011}, Auger and non-Auger carrier-carrier scattering \cite{Li2010,Kim2011,Rana2007}, as well as carrier-phonon scattering with the relevant optical phonon branches \cite{Kampfrath2005,Butscher2007,Tse2009,Kang2010}. Both azimuthal populations are assumed to possess half of the angularly integrated density of states $D_\parallel(\varepsilon)=D_\perp(\varepsilon)=D(\varepsilon)/2=|\varepsilon|/\pi\hbar^2v_F^2$, where $v_F$ is the Fermi velocity \cite{CastroNeto2009}. Particularly, this approach not only provides a realistic description of carrier relaxation, but also gives direct access to the spectral shape of the PL emission. Different from conventional band gap emission in semiconductors, the change of the carrier distribution due to photon emission is negligible in the semimetal graphene and therefore not taken into account. (See Supplemental Material for a detailed description of the numerical approach \cite{Note1}.)

It should be noted that the incoherent PL due to radiative carrier-carrier recombination described in the microscopic modeling may be accompanied by a recently proposed broadband coherent contribution \cite{Winzer2015}, which would be inherently polarized. However, we experimentally verified that the PL detected in our experiment is a fully incoherent emission originating from a non-equilibrium carrier population, and off-diagonal elements of the carrier density matrix can be neglected \cite{Gaida}.

\begin{figure*}[t]
		\includegraphics[width=\textwidth]{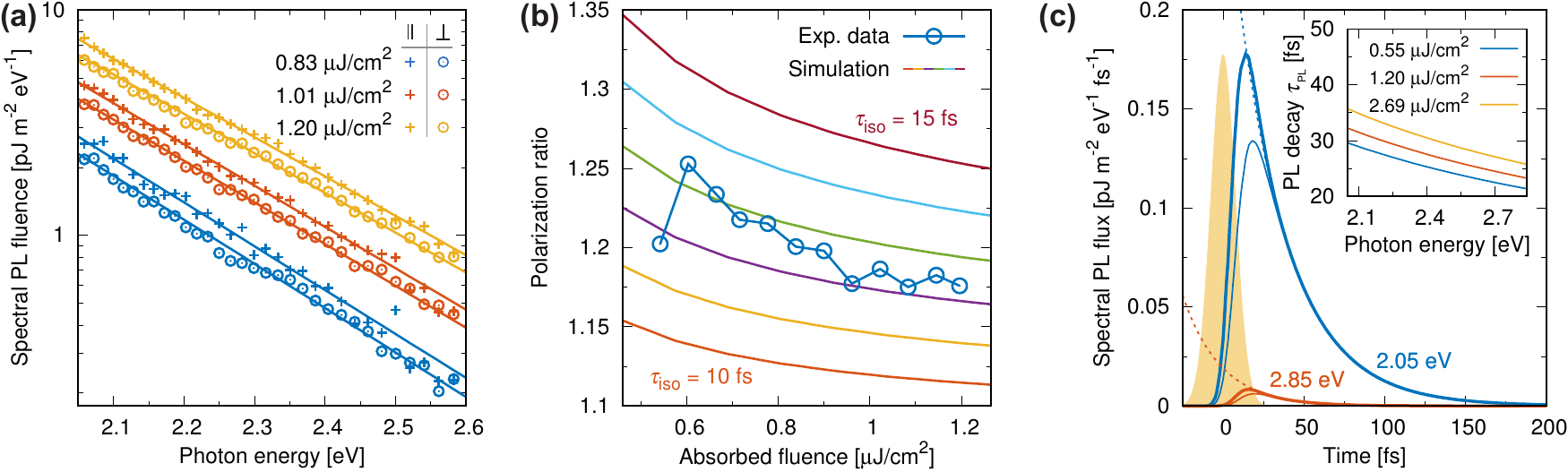}
    \caption{\label{fig2}
		\textbf{(a)} Comparison of experimentally obtained PL spectra (data points) with results of the microscopic simulations (solid lines) ($\tau_\text{iso}=12.5\,\text{fs}$). The absolute PL emission fluence is obtained from the simulations.
		\textbf{(b)} Photon energy-averaged data of the polarization ratio as a function of absorbed fluence. Simulation results are shown for a set of isotropization times between $10\,\text{fs}$ and $15\,\text{fs}$ in comparison with experimental data.
		\textbf{(c)} Simulated PL flux transients at an absorbed fluence of $1.20\,\text{\textmu J}/\text{cm}^2$ for the lowest and the highest PL photon energies accessible in the experiment (thick line: parallel PL, thin line: perpendicular PL, $\tau_\text{iso}=12.5\,\text{fs}$). The temporal shape of the pump pulse is shown for comparison (shaded area). Mono-exponential fits starting from the right-hand inflection point of the emission curves (dashed lines) determine the effective PL lifetime $\tau_\text{\tiny PL}$.
		\textbf{Inset:} Effective PL lifetime $\tau_\text{\tiny PL}$ for different absorbed fluences.
		}
\end{figure*}

Results of the microscopic simulations are depicted in Fig.~\ref{fig2}a (solid lines) in comparison with experimental PL spectra (data points) for three exemplary absorbed fluences. In the simulation, the spectral shape is well reproduced by adopting absorbed fluences of $0.83\,\text{\textmu J}/\text{cm}^2$ to $1.20\,\text{\textmu J}/\text{cm}^2$ for the experimental incident fluence range of $82\,\text{\textmu J}/\text{cm}^2$ to $117\,\text{\textmu J}/\text{cm}^2$. Within the experimental uncertainties, the corresponding graphene absorbance agrees with reported values in the literature \cite{Nair2008,Merthe2016}. The simulated PL spectral flux for an absorbed fluence of $1.20\,\text{\textmu J}/\text{cm}^2$ results in a quantum efficiency of $3\times 10^{-10}$, close to the experimental value of ${\sim}10^{-10}$, which was derived from the PL intensity by taking into account an estimated total detection efficiency of $1\,\text{\%}$ to $2\,\text{\%}$ in the considered photon energy range ($2.05\,\text{eV}$ to $2.85\,\text{eV}$).

Figure~\ref{fig2}b shows a comparison of experimental data for the polarization ratio with simulated curves for a set of isotropization times, each averaged over the experimentally accessible photon energy interval. The time-integrated polarization ratio extracted from the numerical simulations increases with decreasing absorbed fluence and increasing isotropization time. Considering the experimentally observed polarization ratio, we obtain a fluence- and photon energy-averaged value of $\tau_\text{iso}=12\pm 2\,\text{fs}$.
For comparison, in Ref.~\onlinecite{Malic2012}, the authors predict the formation of an isotropic electron distribution within $50\,\text{fs}$ at comparable fluences, while first indications of momentum relaxation are observed after $10\,\text{fs}$. Notably, at optical excitation energies in the meV range, the lifetime of the anisotropic carrier distribution exceeds $1\,\text{ps}$ \cite{Konig-Otto2016}.
 
For comparison with our phenomenological model, we also extract the effective PL lifetime $\tau_\text{\tiny PL}$ from the simulated time-resolved PL emission in a photon energy- and fluence-resolved manner using a mono-exponential model (see Fig.~\ref{fig2}c). It should be noted that $\tau_\text{\tiny PL}$ is dominated by the energetic carrier relaxation timescale $\tau_\text{rel}$, and does not correspond to the intrinsic radiative lifetime of excited carriers in graphene (which is typically on the order of picoseconds \cite{Rana2007,Limmer2013}). Specifically, as the carrier occupation enters quadratically into the PL flux, $\tau_\text{rel}$ is approximately twice as large as $\tau_\text{\tiny PL}$.

Values of $\tau_\text{\tiny PL}$ obtained from the fits in Fig.~\ref{fig2}c are depicted in the inset, exhibiting slower PL decay at low photon energies and for increasing excitation levels, respectively. For the experimental parameters, the estimated $\tau_\text{\tiny PL}$ ranges from $21\,\text{fs}$ to $32\,\text{fs}$, corresponding to an energetic relaxation time constant $\tau_\text{rel}$ between $42\,\text{fs}$ and $64\,\text{fs}$. Thus, we obtain a ratio $\tau_\text{rel}/\tau_\text{iso}$ of 3.5 to 5.3, which is in reasonable agreement with the factor of 8 obtained from the simplified, phenomenological model (cf. Fig.~\ref{fig1}c). Furthermore, the fluence dependence of $\tau_\text{\tiny PL}$ qualitatively accounts for the experimentally observed PL polarization decrease at high fluences, as, in this case, the PL emission contains larger contributions from the isotropic carrier distribution at later times. Besides, from the photon energy dependence of $\tau_\text{\tiny PL}$, a pronounced change of the polarization ratio across the PL spectrum is expected, which, however, is not observed experimentally.

\begin{figure*}[t]
		\includegraphics[width=\textwidth]{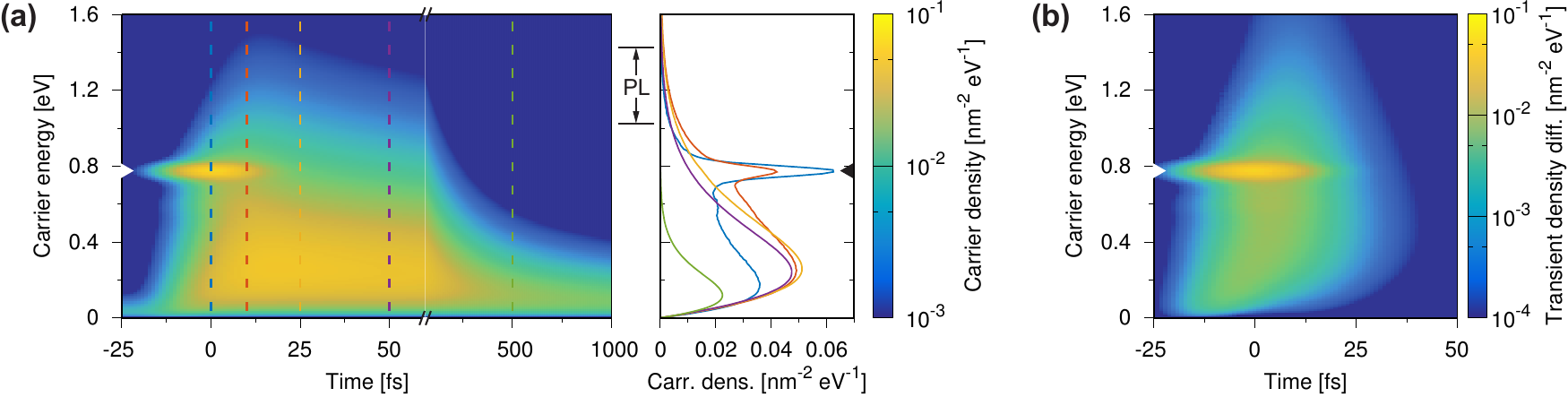}
    \caption{\label{fig3}
		\textbf{(a)} Time-resolved carrier density map of the parallel component $D_\parallel\, f_{c,\parallel}$ at an absorbed fluence of $1.20\,\text{\textmu J}/\text{cm}^2$ ($\tau_\text{iso}=12.5\,\text{fs}$). Lineouts show the carrier distribution at times of 0, 10, 25, 50 and $500\,\text{fs}$. The carrier energy range accessible by PL spectroscopy in our experiment is indicated.
		\textbf{(b)} Transient carrier density difference $D_\parallel\, f_{c,\parallel}-D_\perp\, f_{c,\perp}$ for the same parameters as in (a).
		}
\end{figure*}

To obtain further insights into the temporal evolution of graphene excitation underlying anisotropic PL emission, Fig.~\ref{fig3}a displays the numerically obtained carrier density map $D_\parallel\,f_{c,\parallel}$ of the parallel component of the carrier population within the first picosecond after optical excitation. During pulse absorption, a pronounced occupation builds up at half the pump photon energy (indicated by arrow). Carrier-carrier scattering results in a fast fading of the initial occupation peak and a pronounced broadening of the carrier distribution within the first ten femtoseconds. The emerging high-energy tail is the source of the experimentally observed PL. Slower carrier-phonon-mediated carrier cooling on a $100\,\text{fs}$ timescale \cite{Kampfrath2005,Butscher2007} quenches PL emission. Figure~\ref{fig3}b depicts the difference in carrier densities between the parallel and perpendicular components, illustrating the filling of $D_\perp\, f_{c,\perp}$ due to equienergetic azimuthal scattering. Notably, both energetic broadening and momentum isotropization take place on a timescale comparable to the duration of the pump pulses. Specifically, the transient carrier density difference promptly increases at early delay times, and the carrier anisotropy has almost vanished again after $25\,\text{fs}$.

Finally, we want to point out the connection between the experimentally observed PL polarization ratio $R$ and the quantum mechanical pseudospin degrees of freedom in graphene. Considering graphene's optical matrix elements \cite{Mecklenburg2010}, one arrives at the following connection between $R$ and the expectation values of the sublattice pseudospin in $q_x$ and $q_y$ directions:
\begin{equation}
    R(E)=\frac{\left[\left|\left<q|\hat\sigma_x|q\right>\right|^2\right]}{\left[\left|\left<q|\hat\sigma_y|q\right>\right|^2\right]}.
\end{equation}
Here, $\hat\sigma_x$ and $\hat\sigma_y$ are the pseudospin operators and $\left|q\right>$ is the electron state with momentum $\bm q$ (relative to the $K$ and $K'$ point). $[\cdot]$ denotes the temporal average over the occupied conduction electron states at energy $\varepsilon(\bm q)=E/2$, i.\,e. $[\cdot]=\int \text{d}t \sum_{\bm q}\,f_c^2\,\delta(\varepsilon-E/2)$. Thereby, the degree of polarization in the PL spectrum in graphene gives a direct fingerprint of the non-equilibrium pseudospin alignment on ultrashort timescales.

In summary, we presented polarization-resolved PL spectroscopy as a robust tool to directly access anisotropic carrier distributions in graphene during the first $10\,\text{fs}$ to $20\,\text{fs}$ after ultrashort optical excitation. In combination with a microscopic Boltzmann simulation, we extract a characteristic isotropization timescale of $12\pm 2\,\text{fs}$, resulting in a rapid loss of initial pseudospin alignment imprinted by the optical field.

\begin{acknowledgments}
The authors thank H. K. Yu and M. Maiti for CVD preparation, and S. Dechert for Raman characterization of the examined graphene samples. PMMA spin-coating and reactive-ion etching were conducted with support by S. Schweda and M. Sivis, respectively. Partial funding was provided by DFG-SFB-1073 `Atomic Scale Control of Energy Conversion', project A05. TD gratefully acknowledges a scholarship by the German Academic Scholarship Foundation. 
\end{acknowledgments}

\end{document}